# Fallen Angel Bonds Investment and Bankruptcy Predictions Using Manual Models and Automated Machine Learning

Harrison Mateika, Juannan Jia, Linda Lillard, Noah Cronbaugh, and Will Shin

**Northwestern University**
HarrisonMateika2022@u.northwestern.edu
JuannanJia2020@u.northwestern.edu
LindaLillard2022@u.northwestern.edu
NoahCronbaugh2022@u.northwestern.edu
WilliamShin2022@u.northwestern.edu

December 4, 2022

The primary aim of this research was to find a model that best predicts which fallen angel bonds would either potentially rise up back to investment grade bonds and which ones would fall into bankruptcy. To implement the solution, we thought that the ideal method would be to create an optimal machine learning model that could predict bankruptcies. Among the many machine learning models out there we decided to pick four classification methods: logistic regression, k-nearest neighbor, support vector machines, and neural networks. We also utilized an automated methods of machine learning: Google Cloud's AutoML.

The results of our model comparisons showed that the models did not predict bankruptcies very well on the original data set with the exception of AutoML having a high precision score. However, our oversampled and feature selection data set did perform very well. However, this could likely be due to the model being overfitted to match the narrative of the oversampled data (as in, it doesn't accurately predict data outside of this data set quite well). Therefore, we were not able to create a model that we are confident that would predict bankruptcies.

However, we were able to find value out of this project in two key ways. The first is that the AutoML model in every metric and in every data set either outperformed or performed on par with the other models. The second is that we found that utilizing feature selection did not reduce predictive power that much. This means that we can reduce the amount of data to collect for future experimentation regarding predicting bankruptcies.

## 1. Introduction

The goal of investing in bonds is to maximize the return on an investment to the highest degree possible. This can often mean the blending together of two diametrically opposed methods: investing in conservative bonds that are guaranteed a return, or investing in aggressive bonds that carry a substantially increased amount of risk. The conservative bonds tend to guarantee a return, but the return is fairly low due to its low yield (also known as the interest an investor receives on a bond). However, the riskier bonds have a higher yield and thus can result in a far greater return. Due to this, investors often have to make the unenviable decision to invest in bonds that they know are at a higher risk of failure.

The bond market is in a unique position when compared to other financial markets due to its rating system. This rating system is oligarchical by nature, since the ratings are given out by three bond rating agencies. All of these agencies follow the exact same ratings system: AAA is the highest, while D is the lowest. Bonds that are highly rated, and thus conservative, are referred to as investment-grade bonds. These bonds typically fall in the AAA, AA, A, or BBB ratings. Bonds that are rated lower than that are referred to as non-investment-grade bonds. These bonds are also unceremoniously known as junk bonds.

Junk bonds usually consist of two types of companies: small startups or companies with high debt ratios. These are normally the types of risky bonds that investors struggle to invest in. For this reason, they are often looking for an opportunity to mitigate the risk of failure. One potential opportunity is with Fallen Angel and Rising Star bonds. A Fallen Angel refers to a bond that once had an investment-grade rating, but fell to junk bond territory due to its company's financial hardship. A Rising Star refers to a bond that was once considered a junk bond, but rose to an investment-grade bond.

The essential idea is that if an investor invests in a Fallen Angel, the investor can take advantage of the bond's currently high yield and low purchase price. If the bond were to become a Rising Star, the bond's price would rise, and allow the investor to sell it for a profit. However, the investor also is taking on a major risk. Since a Fallen Angel is often a company that has fallen on hard times, it also is likely that the bond could continue to fall into bankruptcy. Fallen Angels and junk bonds are everywhere and often one of the challenges associated with them is understanding the inherent risk of investing in these bonds. While the rating system is a helpful barometer, the amount of information it gives regarding whether or not a company could go bankrupt is dubious.

Further complicating the matter is the amount of information that goes into determining how a company is performing and what direction it's going in. The amount of information is far too much for an individual to take in, and thus requires methods beyond financial ratios and eye-scans. The challenge of determining the chance of a company going bankrupt, therefore, deserves a degree of scrutiny that cannot be accomplished using typical methods.

We would like to find the optimal machine learning method that can best predict whether or not a company will go bankrupt by running various models manually, and then comparing them not only against each other but also against an automated machine learning tool called AutoML, a product on Google Cloud. This program sets out to implement feature engineering, dimensional reduction, and model choice to ease the process of machine learning.

## 2. Literature Review

Predicting corporate bankruptcy using financial ratios has been something that many data scientists have been looking to do with statistical models for a long period of time. One of the most famous papers in that instance was Edward Altman's paper titled "Financial Ratios, Discriminant Analysis and the Prediction of Corporate Bankruptcy", a paper that came out in 1968. In that paper, Altman utilized multiple discriminant analysis on 22 financial ratios of 66 companies, half of which filed for bankruptcy. The results were that only 5 variables were needed to accurately predict bankruptcy. Utilizing the model, Altman was able to predict bankruptcy for companies very accurately for their first year. However, that accuracy waned as the years went forward:

Much of the work in predicting the bankruptcy of companies has largely been built off of this work. However, regarding machine learning terms, the challenge for many of these papers is that it is difficult to find data with many bankruptcy occurrences. The main reason for this is simple: only 3% of public companies file for bankruptcy. With such a large imbalance between instances, it is going to be challenging to build a precise model.

This challenge was an issue for Vikram Devatha in his article entitled "Predicting bankruptcy using Machine Learning". In his experiment, Devatha utilized several classifications algorithms to help predict bankruptcy including Logistic Regression, Perceptron as a classifier, Deep Neural Network Classifiers, Fischer Linear Discriminant Analysis, K Nearest Neighbor Classifier, Naive Bayes Classifier, Decision Tree, Bagged Decision Trees, Random Forest, Gradient Boosting and Support Vector Machine. He also expanded the concepts by using different initializations. For instance, the author used different k values from 1 - 19 to compare the results for K Nearest Neighbor, and he used a range of 50-500 tree increments for his Random Forest method. His results found that Gradient Boosting and Bagged Decision Trees performed best among the other algorithms when using the Sensitivity (True Positive) measurement as his final result. In this instance, his sensitivity analysis performed rather well, with a True Positive of 85.05%.

Kou, Xu, Peng et. al explored the imbalance of bankruptcies when evaluating the models for their paper Bankruptcy Prediction for SMEs using transactional data and two-stage multiobjective feature selection. In the paper, the team compared the model results of their original bankruptcy data set to an undersample and oversample of the data set. The results were that the models performed better on the imbalanced data set and that the sensitivity of the imbalanced data set did not improve when utilizing the sampling techniques. The paper also found that the ensemble method XGB performed best when compared to the other models when looking at AUC. Model 5 in this paper was reflective of transactional data being included in predicting bankruptcies, giving this paper some real practical implications.

Liang, Lu, Tsai, and Shih for their paper Financial ratios and corporate governance indicators in bankruptcy prediction: A comprehensive study looked at the same data set that we are utilizing for this paper. Their primary aim was to look at the corporate governance ratios associated with this data and determine whether or not it increased the data's predictive power. For their results, the corporate governance ratios did prove to increase the predictive power of determining bankruptcies. When looking at the statistical techniques they utilized (support vector machines, k-nearest neighbor, naive Bayes classifier, classification and regression tree, and multilayer perception methods), they found that SVMs performed the best.

Regarding which papers performed the best or worst, all of these papers struggle with the same challenge: a large data imbalance regarding what companies went into bankruptcy and those that did not. Each of these papers tackles the approach in a different way, and each has a multitude of implications. Altman's paper was a relic of its time, and while still highly relevant, it lacks the sophisticated techniques that we have today to determine bankruptcy. Furthermore, his paper readily admits that it lacks practical implications. The papers regarding Kou, Xu, Pen et. al and Liang, Lu, Tsai et.al each have their own practical implications associated with them: for Kou, Xu, Pen et. al the implication was mainly to do with transactional data while for Liang, Lu, Tsai et.al it had to do with corporate governance data. Due to both having real-world implications, we have decided that both deserve the most recognition as the most rigorous analysis of bankruptcy data.

Regarding our contribution to bankruptcy data literature, we differentiate ourselves from these papers in a few ways. First, we are oversampling the corporate governance data used by Liang, Lu, Tsai et.al to determine whether or not it outperforms the data imbalance. This is mainly to see if the results from Kou, Xu, Pen et. al is replicable with other bankruptcy data. Second, we are implementing rigorous analysis to our methods. While we are looking to determine the best model utilizing manual feature selection, modeling, and evaluation methods, we are also determining whether or not our methods outperform automatic machine learning methods through AutoML. This determination will mean that there will be practical implications to our paper no matter what: either we can outperform automated methods in determining bankruptcies, or perhaps automated machine learning methods are the best method of dealing with this type of data.

## 3. Data Collection

The data was collected from Kaggle data sets. It contained 10 years of Taiwanese company data with financial information for 6,819 companies between the years 1999 to 2009, published by the Taiwan Economic Journal. The data set includes 95 financial data ratios, ratios regarding corporate governance, and a bankruptcy indicator (1=bankrupt, 0=non-bankrupt). Corporate bankruptcy was defined by the business rules of the Taiwan Stock Exchange (Wang and Liu 2021). We found no non-null values, and the data types were either integers (int64) or floats (float64).

Upon the initial exploratory data analysis, we uncovered that, like most data containing bankruptcy data, the data contained in the Taiwanese company bankruptcy data set is highly skewed towards financially stable (non-bankrupt) companies. Training our models using the data set will likely produce overfitting and biased results toward non-bankrupt companies.

As seen on the graph, there is a large data imbalance between bankrupt and non-bankrupt companies in this data set. This will pose a problem as most machine learning algorithms will not work well with imbalanced data sets. The training model would not learn enough about the bankrupt company data if proper sampling techniques are not applied.

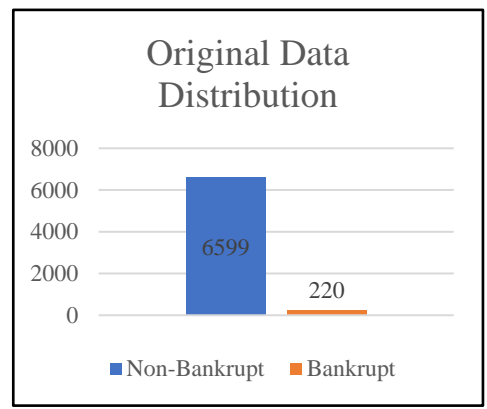

To address the data set imbalance, we created a separate data set representing more balanced data across bankrupt and non-bankrupt companies. We used SMOTE (Synthetic Minority Oversampling Technique) to create an oversampling data set. SMOTE works by selecting data close to the feature space, drawing a line between the examples in the feature space, and drawing a new sample at the point along that line. (Brownlee 2020) We used the oversample.fit_resample() method to create a new oversampled data file containing 26,396 perfectly balanced records with 13,198 bankrupt data sets and 13,198 non-bankrupt data sets.

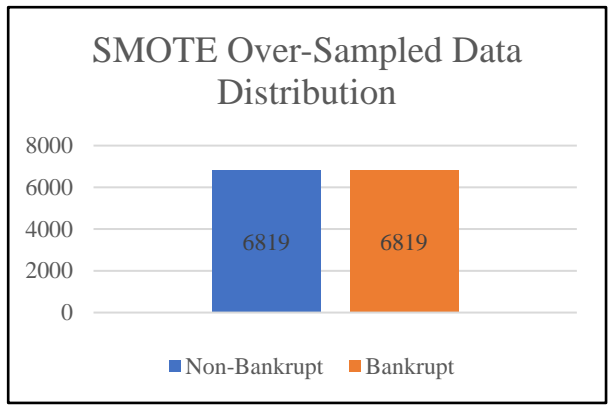

### 4. Data Analysis.

During the exploratory data analysis (EDA) stage, we used a correlation heatmap matrix to identify the ten highest correlated features to bankruptcy. Out of the ten, we found the five commonly high features correlated with bankruptcy to be Debt Ratio Percentage, Current Liability to Assets, Current Liability to Current Assets, Total Expense / Assets, and Cash / Current Liability. We then compared both the top positive and negative correlated features, we found that organizations that possess more assets and earnings are healthier and less likely to be bankrupt.

Other observations during the EDA process included:
- Most of the features have outliers. Median will be a better analysis method and, also, taking some outliers out will be a good idea when building the model.
- Companies with a low 'Net profit before tax/Paid-in capital', 'Persistent EPS', and 'Net Value Per Share (A)' tend to go bankrupt.
- 'Borrowing dependency' has bankrupt companies distributed throughout all its range. However, around 0.4 are located in the companies that do not go bankrupt. Having around 0.4 does not guarantee to be bankrupt safe since many companies went bankrupt with this index, but having a higher o lower index seems critical since there are not any companies operating with this kind of index.
- 0.8 "Net Income to Stockholder's Equity" is an excellent indicator to operate but does not entirely save you from bankruptcy.
- The number of organizations that have gone bankrupt in 10 years between 1999 – 2009 is few.
- Very few organizations with negative income have suffered from bankruptcy in the past two years.
- An increase in the values of the attributes that negatively correlate with the target attribute helps an organization avoid bankruptcy.

### 5. Methodology

We split the data set into 70/30 for training and testing. Different classifier algorithms were explored to build an optimal model.

To measure the success of a model, the confusion matrix was used to identify true positives, true negatives, false positives, and false negatives.

We gave heavy weight to true positives and true negatives since those metrics indicate the number of bankruptcies determined accurately and the number of non-bankruptcies determined accurately.

*5.1 Logistic Regression (LR)*

In the LR model performed on the original data set, 99% correctly labeled non-bankruptcy data as nonbankruptcy, and 16% correctly labeled the actual bankruptcy data as bankruptcy. The high degree of accuracy (99%) to predict the true non-bankruptcy event and the low degree of accuracy (16%) to predict the true bankruptcy event show the bias of how much non-bankruptcy data is in the set. The original data set's precision, recall, f1-score, and accuracy were 0.35, 0.16, and 0.22, respectively. This indicates that the model does a mediocre job of predicting bankruptcies.

In the LR model performed on the oversampled data set, 88% correctly labeled the actual non-bankruptcy data as the non-bankruptcy and 92% correctly labeled the actual bankruptcy data as bankruptcy, which showed the balanced prediction for the true positives and true negatives to avoid the bias produced in the first data set. The precision, recall, and f1-score were 0.89, 0.92, and 0.90. This is, of course, a massive improvement over the previous data set but can largely be attributed to the fact that the model may potentially be overfitted.

In the LR model that performed the oversampled data set with the feature selection, 89% correctly labeled the actual non-bankruptcy data as the non-bankruptcy, and 90% correctly labeled the actual bankruptcy data as a bankruptcy. The precision, recall, and f-score are 0.88, 0.89, and 0.89, respectively. Compared with the performance data from the previous data set, the feature selection did not improve the model performance significantly.

*5.2 K-Nearest Neighbors (KNN)*

In the KNN model that performed on the original data set, 100% correctly labeled the actual non-bankruptcy data as the non-bankruptcy and 12% correctly labeled the actual bankruptcy data as bankruptcy. The high degree of accuracy (almost 100%) to predict the true non-bankruptcy event and low degree of accuracy (12%) to predict the true bankruptcy event raised the red flag of the bias issue again, like the Logistic Regression model. The original data set's precision, recall, and f1-score are 0.48, 0.16, and 0.24, respectively. In terms of the aggregate metric (the f1 score) this model did not perform as well as the logistic regression on the original data set. However, it did have a slightly higher precision score.

In the KNN model that performed on the oversampled data set, 88% correctly labeled the actual nonbankruptcy data as non-bankruptcy, and 100% correctly labeled the actual bankruptcy as bankruptcy. The precision, recall, and f1-score for the oversampled data were 0.89, 1, and 0.94, respectively. Like the logistic regression model, the oversampled data outperformed the original data set. Furthermore, this model outperformed the logistic regression in almost every metric.

In the KNN model that performed on the oversampled data set with the feature selection, 89% correctly labeled the actual non-bankruptcy data as non-bankruptcy, and 96.6% correctly labeled the actual bankruptcy as bankruptcy. The data set's precision, recall, and f1-score were 0.88, 0.94, and 0.9, respectively. Compared with the performance data from the previous data set, the model with the feature selection generated similar prediction results and did not improve the model performance significantly. This is similar to both of the previous model results.

*5.3 Support Vector Model (SVM)*

In the SVM model that performed on the original data set, 99.9% correctly labeled the actual non-bankruptcy data as the non-bankruptcy, and 1.5% correctly labeled the actual bankruptcy as the bankruptcy. The high degree of accuracy (almost 100%) to predict the true non-bankruptcy event and the low degree of accuracy (1.5%) to predict the true bankruptcy event detected the bias issue. The original data set's precision, recall, and f-score are 0.33, 0.02, and 0.04. This model performed far worse than the previous two models, mainly due to its recall score.

In the SVM model that performed on the oversampled data set, 92% correctly labeled the actual non-bankruptcy data as the non-bankruptcy, and 100% correctly labeled the actual bankruptcy as bankruptcy, which showed the balanced prediction for the true positives and true negatives. The precision, recall, and f-score for the oversampled data set are 0.93, 0.98, and 0.95, respectively. This version of the model outperformed the logistic regression and the KNN model on all metrics.

This is fascinating since the original data set for this performed worse than the other two. Perhaps this model performs better as a whole with more information compared to the other two.

In the SVM model that performed on the oversampled data set with the feature selection, 88% correctly labeled the actual non-bankruptcy data as the non-bankruptcy, and 92.8% correctly labeled the actual non-bankruptcy as the non-bankruptcy. The data set's precision, recall, and f-score are 0.88, 0.94, and 0.90, respectively. The model on this data set had an inferior performance compared to the other oversampled data set, following similar trends from the other two models.

*5.4 Neural Network (NN) Model*

We used Keras to perform a multi-layer feed-forward neural network model. The first hidden layer has 16 nodes and uses the "relu" activation function. The second hidden layer has 16 nodes and uses the relu activation function. The output layer has 1 node and uses the sigmoid activation function. Since our model is a binary classification problem, "binary_crossentropy" is used as the loss argument. The "epoch" value used was 120 with a batch_size of 10.

In the NN model that performed on the original data set, 99.9% correctly labeled the actual non-bankruptcy data as non-bankruptcy, and 1.9% correctly labeled the actual bankruptcy as a bankruptcy. Once again, the high degree of accuracy (almost 100%) to predict the true non-bankruptcy event and the low degree of accuracy (1.9%) to predict the true bankruptcy event generated skewness toward the non-bankruptcy data. The original data set's precision, recall, and f-score are 0.33, 0.02, and 0.04, respectively. This model performed just as well as the SVM model on most metrics.

In the NN model that performed on the oversampled data set, 97% correctly labeled the actual non-bankruptcy data as non-bankruptcy, and 99% correctly labeled the actual bankruptcy as bankruptcy. The precision, recall, and f-score for the oversampled data set are 0.97, 0.99, and 0.98, respectively. This model on the oversampled data set is the best model regarding performance and has the highest precision, recall, and f score as a whole.

In the NN model that performed on the oversampled data set with the feature selection, 92.9% correctly labeled the actual non-bankruptcy data as the non-bankruptcy and 95.9% correctly labeled the actual non-bankruptcy as the non-bankruptcy. The precision, recall, and f scores were 0.93, 0.96, and 0.94, respectively. This model performed worse than how it performed on the oversampled data set. However, this model outperformed the other models regarding this data set.

*5.5 Auto Machine Learning (AutoML)*

Within the original data set, the AutoML model overall produced a model that may be considered the best or second best depending on how one looks at the results. If the aim is to predict the True Positive, the model AutoML performed second to the logistic regression model (.14 vs .15). However, if one looks at the precision of bankruptcies predicted, the model performed by far the best at a precision of .60. When looking at an aggregate metric though like the f score, the model performed the third best at 0.22. This is primarily due to the model's lower recall score when compared to logistic and KNN. Overall, the model as a whole is fairly comparable to the other models and its precision rate is definitely not something to be overlooked.

When looking at the oversample results, the AutoML performed the best out of the rest of the models for most of the metrics. The AutoML model had the highest precision (0.98), is tied with KNN for the highest recall (1) and the highest f-score (0.99). The only metric where the AutoML is outperformed is on the True Negative where it is outperformed by the neural network model with AutoML having 0.98 and the neural network model having 0.99. Overall, since AutoML outperformed the other models regarding metrics, it may be the best performer.

When looking at the oversample with feature selection results, the AutoML model performed the best regarding every metric when compared to the other models. It had the highest precision (0.95), recall (1) and f score (0.98). Like the other models those scores were not as high as the score with all of the features included.

## 6. Results

Regarding the actual results of predicting bankruptcies across all models, there was a recurring pattern of the original data set produced low scoring metrics across all models when it came to accurately predicting bankruptcies (this was especially the case with the recall scores and the true positive scores). The oversampled data set produced high scoring metrics across all models. The oversampled data set with feature selection would produce high scoring metrics but not quite as high as the normal oversampled data set. Finally, among all three of these data sets, one of the highest scoring models was the model generated by AutoML.

The reason that the original data set performed poorly included the data set imbalance between bankrupt companies and non-bankrupt companies. However, within the original data set, the AutoML model outperformed the other models.

Both oversampled data sets performed better than the original data set. However, this high degree of predictability comes at a cost: the model is very likely to be overfitted. The reason for this is that to oversample we had to repeat the bankruptcy values and associated rows.

## 7. Analysis and Interpretation

Due to the complicating factors regarding bankruptcy, it is very unlikely that a model will be generated that will be able to accurately predict all bankruptcies. However, the high precision value that the AutoML model generated gives us hope that an accurate model can be created.

## 8. Conclusions and Next Steps

We recommend more bankruptcy data be collected, particularly more recent data. Furthermore, AutoML may be a more cost-saving method of modeling. It allows users to implement machine learning processes with greater ease than running different models with associated parameters manually.

The AutoML process is fairly simple: upload the CSV data or connect to a BigQuery table, pick a target column for prediction, select a metric to optimize, and select an amount of training hours to run the model.

The 2020-2021 COVID pandemic had an unseen impact on the fallen-angel bond market. Credit rating agencies have downgraded many companies that were deemed at risk of being unable to repay investors when their bonds mature. By the end of 2020, the fallen-angel universe more than doubled compared to pre-pandemic. Many of these bonds have outperformed the overall high-yield market and investment-grade bonds in the last decade. Buying fallen-angel bonds in the clients' portfolios can yield a significant return when these companies improve their finances and upgrade back to investment grade. When this occurs, the prices will spike when these bonds are again eligible for investment in the investment group portfolios. In the first half of 2022, about $69 billion of high-yield bonds were upgraded to investment grade (Veraa, 2022).

This information above represents a great opportunity to take the research to the next level. By collecting the data on these fallen angel bonds, we would be able to gain a higher set of data regarding companies that go bankrupt than ever before. This should be an area to direct the next data science team toward regarding data collection.